\documentclass[11pt,twoside,onecolumn]{article}
\usepackage[]{latexsym}
\usepackage{epsfig}
\usepackage{amsmath,amssymb}
\newcommand{\be}{\begin{eqnarray}}
\newcommand{\ee}{\end{eqnarray}}
\newcommand{\ud}{\mathrm{d}}

\newcommand{\lp}{\ell_{\rm p}}
\newcommand{\mpl}{m_{\rm p}}

\title{\bf Charged shells and elementary particles}
\author{Roberto~Casadio$^{a,b}$\thanks{roberto.casadio@bo.infn.it}
\\
\null
\\
$^a${\em Dipartimento di Fisica e Astronomia, Universit\`a di Bologna}
\\
{\em via Irnerio~46, 40126 Bologna, Italy}
\\
\\
$^c${\em Istituto Nazionale di Fisica Nucleare, Sezione di Bologna}
\\
{\em via Irnerio~46, 40126 Bologna, Italy}
}
\begin{document}
\maketitle
\begin{abstract}
We review the General Relativistic model of a (quasi) point-like particle represented by
a massive shell of electrically charged matter, which displays an ADM mass $M$ equal to the
electric charge $|Q|$ in the small-volume limit.
We employ the Israel-Darboux's junction equations to explicitly derive this result, and then
study the modifications introduced by the existence of a minimum length scale $\lambda$.
For $\lambda$ of the order of the Planck length (or larger), we find that the ADM mass
becomes equal to the bare mass $m_0$ of the shell, like it occurs for the neutral case.  
\end{abstract}
\section{Introduction and general perspective}
\label{secIntro}
\setcounter{equation}{0}
Elementary particles are usually viewed as being point-like in classical physics,
although the stress-energy tensor of the electromagnetic and Newtonian gravitational
fields then diverge at the particle's location.
This divergence can be removed in General Relativity,
by replacing localised sources with shells of matter~\cite{adm60},
whose ``size'' and total energy, the Arnowitt-Deser-Misner (ADM)
mass $M$, remain finite in the ``point-like'' limit.
This result is further interpreted as the fact that General Relativity
does not allow to store finite amounts of energy in a vanishingly small volume.
\par
In Quantum Mechanics, the Heisenberg Uncertainty Principle prevents complete
localisation in the phase space of Minkowskian theories.
Moving forward to a semiclassical scenario, with quantum matter evolving on 
a classical background space-time, rigorous results and plausibility arguments
suggest the emergence of a fundamental length scale, say
$\lambda$~\cite{gup,string,loop,szabo}.
Such a scale is usually predicted to be of the order of the Planck length
$\lp\simeq 10^{-33}\,$cm, corresponding to a mass $\mpl\simeq 10^{16}\,$TeV,
well beyond the realm of earth-based experiments.
From a theoretical point of view, it is still interesting to explore the conceptual
implications of the existence of a length $\lambda\sim\lp$ on the fundamental nature
of elementary particles. 
More specific questions then are, for example, if black holes differ significantly
from standard model particles and whether quantum transitions may occur between
black holes and regular particles at the Planck scale.
The latter issue is particularly relevant for the understanding of the end-point of
Hawking evaporation, but might also be relevant for studying the formation
of horizons inside collapsing matter.
\par
We shall first re-derive the results of Ref.~\cite{adm60} by making use of
Israel-Darboux's junction equations~\cite{israel}, showing that (spinless)
electrically charged particles with bare mass $m_0$ can be described in
General Relativity by extremal configurations of the Reissner-Nordstroem
metric with (geometrical) charge $|Q|=M$.
We shall then introduce a minimum length $\lambda\gtrsim\lp$, thus extending
the analysis previously performed for the neutral case~\cite{cgs},
and find that the main result of Ref.~\cite{cgs}, namely that $M\simeq m_0$,
still holds for $Q\not=0$.
Some connections with recent models of the internal degrees of freedom
of black holes will also be mentioned.
\par
For simplicity, we shall mostly use units with $G=c=1$, so that all variables have dimension
of length to a given power, unless differently specified.
\section{ADM shell model}
\label{secADM}
\setcounter{equation}{0}
Following Ref.~\cite{adm60}, we consider the space-time generated by an
infinitely thin shell of bare mass $m_0$, electric charge $Q$ and
(isotropic coordinate~\footnote{The precise definition of $r$ will be given
below in Section~\ref{subsecExt}.})
radius $r=\epsilon$.
Space will therefore be divided into an interior region ($0\le r<\epsilon$) and
exterior region ($r>\epsilon$), relative to the shell radius $r=\epsilon$.
\subsection{Interior geometry}
For the interior region ($0\le r<\epsilon$), we shall assume flat Minkowski space-time,
which, in the usual spherical coordinates with areal radius $\bar r$, of course, reads
\be
\ud s^2_{\rm o}
=
-f_{\rm i}\,\ud t^2_{\rm i}+f_{\rm i}^{-1}\,\ud \bar r^2_{\rm i}+\bar r^2_{\rm i}\,\ud\Omega^2
\ ,
\label{flatg}
\ee
with $f_{\rm i}=1$.
\subsection{Exterior geometry}
\label{subsecExt}
We assume the exterior region ($r>\epsilon$) is described by the Reissner-Nordstr\"om metric,
which is written, in Schwarzschild-like coordinates, as 
\be
\ud s^2_{\rm o}
=
-f_{\rm o}\,\ud t^2+f_{\rm o}^{-1}\,\ud \bar r^2+\bar r^2\,\ud\Omega^2
\ ,
\label{RN}
\ee
where
\be
f_{\rm o}=1-\frac{2\,M}{\bar r}+\frac{Q^2}{\bar r^2}
\ ,
\ee
and the constant $M$ is the ADM mass of the system, as we shall show below.
Let us recall the above metric displays (up to) two horizons
$\bar r=\bar R_\pm$, namely
\be
\bar R_\pm
=
M\pm\sqrt{M^2-Q^2}
\ ,
\label{Rh}
\ee
provided the charge is small enough, that is $|Q|\le M$.
For $|Q|>M$, the metric~\eqref{RN} instead represents a naked singularity located
at $\bar r=0$ (if that region of space is accessible).
In our case, since the outer geometry ends at $\bar r=\bar r(\epsilon)>0$, where
the shell is located, the singularity is not part of the physical space-time.
Nonetheless, in the following we shall pay particular attention to those to cases,
namely $|Q|\le M$ and $|Q|>M$.
\par
\begin{figure}[t]
\centering
\raisebox{3.5cm}{$\frac{\bar r}{M}$}
\epsfxsize=7cm
\epsfbox{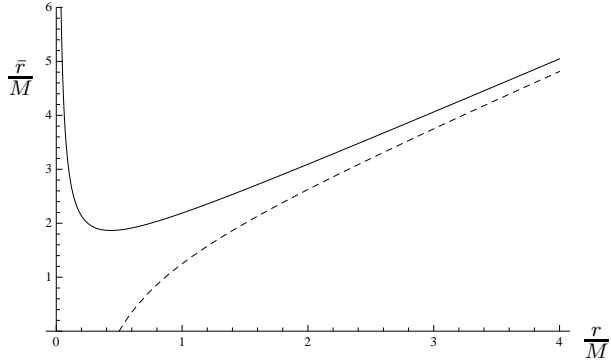}
$\frac{r}{M}$
\caption{Radial coordinate $\bar r$ as function of isotropic radius $r$
for fixed $M$ and $|Q|=M/2<M$ (solid line):
the throat is at $r_+/M=\sqrt{3}/4\simeq 0.43$ with
areal radius $\bar R_+/M=(2+\sqrt{3})/2\simeq 1.86$.
Same function for $M$ and $|Q|=2\,M>M$ (dashed line):
there is no throat and $\bar r(r_{\rm min})=0$ for $r_{\rm min}/M=1/2$.
\label{barr}
}
\end{figure}
%
%
%
%
Before we proceed, we provide the isotropic form of the metric,
namely
\be
\ud s^2_{\rm o}
=
-\left(\frac{4\,r^2-M^2+Q^2}{4\,r^2+4\,M\,r+M^2-Q^2}\right)^2
\ud t^2
+\left(1+\frac{M-Q}{2\,r}\right)^2
\left(1+\frac{M+Q}{2\,r}\right)^2
\left(\ud r^2+r^2\,\ud\Omega^2\right)
\ ,
\label{isoS}
\ee
which is obtained from Eq.~\eqref{RN} with the change of radial coordinate 
\be
\bar r(r)
=
r\left(1+\frac{M}{2\,r}\right)^2
-\frac{Q^2}{4\,r}
=
r+M+\frac{M^2-Q^2}{4\,r}
\ .
\label{barrr} 
\ee
From this form, assuming $|Q|< M$, we see that the metric~\eqref{isoS}
can be used to represent a wormhole,  asymptotically flat both for $r\to\infty$ and $r\to 0$.
In fact, if $|Q|< M$, the areal radius $\bar r$ diverges both for $r\to \infty$
and $r\to 0$, and has a minimum for 
\be
r_+
=
\frac{1}{2}\,\sqrt{M^2-Q^2}
\ ,
\ee
corresponding to $\bar r(r_+) =\bar R_+$ (see solid line in Fig.~\ref{barr}).
In this case, the shell can be placed at any $r=\epsilon\ge 0$.
For $|Q|> M$, there is no throat ($\bar R_+$ becomes imaginary) and the areal radius instead
vanishes for finite $r=r_{\rm min}$ (see dashed line in Fig.~\ref{barr}), $\bar r(r_{\rm min})=0$,
where 
\be
r_{\rm min}
=
\frac{1}{2}\left(|Q|-M\right)
>0
\ .
\label{rmin}
\ee
The metric~\eqref{isoS} now does not represent a worm-hole and the shell can only have
an isotropic radius $\epsilon\ge r_{\rm min}$.
Obviously, this complication does not arise with a neutral shell, which can always be
associated to a worm-hole~\cite{cgs}. 
Finally, for $|Q|=M$, the Reissner-Nordstroem metric~\eqref{RN} becomes extremal,
with $\bar R_-=\bar R_+$, and Eq.~\eqref{isoS} reduces to
\be
\ud s^2_{\rm o}
=
-\left(1+\frac{M}{r}\right)^{-2}
\ud t^2
+
\left(1+\frac{M}{r}\right)^2
\left(\ud r^2+r^2\,\ud\Omega^2\right)
\ ,
\label{isoSex}
\ee
with $\bar r(r)=r+M$ and $r\ge r_{\rm min}=-M<0$.
Note that the latter minimum value of $r$ cannot be smoothly reached by taking
$|Q|\to M$, with $|Q|>M$, in Eq.~\eqref{rmin}. 
Extremal configurations will appear to play a decisive role in Section~\ref{secQM}.
\par
\begin{figure}[t]
\centering
\raisebox{3.5cm}{$\frac{y}{M}$}
\epsfxsize=7cm
\epsfbox{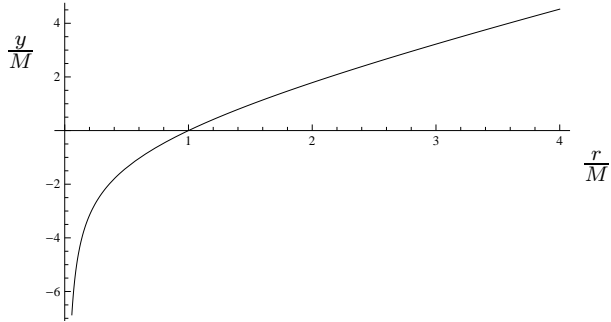}
\raisebox{2cm}{$\frac{r}{M}$}
\caption{Gaussian normal coordinate $y$ as function of isotropic radius $r$
for fixed $M$, $|Q|=M/2$ and $y=0$ for $r=M$.
\label{yvr}
}
\end{figure}
%
%
Another useful parameterization is given by Gaussian normal coordinates,
in which the spatial part of the metric looks flat along the direction perpendicular
to surfaces of constant $r$ or $\bar r=\bar r(r)$ (the usual two-spheres), that is $g_{yy}=1$.
For the metric~\eqref{RN}, the coordinate $y$ is simply defined by the equation  
\be
\ud y
=\frac{\ud\bar r}{\sqrt{f_{\rm o}(\bar r)}}
=\left(\frac{\ud \bar r}{\ud r}\right)
\frac{\ud r}{\sqrt{f_{\rm o}(\bar r(r))}}
\ ,
\ee
which, on using Eq.~\eqref{barrr}, yields
\be
y(r)
=
y_0+r+M\,\ln(r)+\frac{Q^2-M^2}{4\,r}
\ ,
\label{y_r}
\ee
where $y_0$ is an integration constant that can be so chosen that, for example,
$y=0$ for $r=\epsilon$ (see Fig.~\ref{yvr}).
It is now easy to compute the components $K_{ij}$ of the extrinsic curvature
of a two-sphere of isotropic radius $r=x$.
In particular, we shall just need~\footnote{The constant $y_0$ in Eq.~\eqref{y_r}
does not affect the following result and need not be specified.}
\be
K_{\theta\theta}
=
\frac{1}{2}\left.\!\frac{\ud\bar r^2}{\ud y}\right|_{r=x}
=
\bar r(x)\,\sqrt{f_{\rm o}(\bar r(x))}
\ ,
\ee
from which we also obtain the trace $K=K^i_{\ i}$,
\be
K
=
2\,g^{\theta\theta}\,K_{\theta\theta}
=
2\,\frac{\sqrt{f_{\rm o}(\bar r)}}{\bar r}
\ .
\label{K}
\ee
Note that, for $Q=0$, we recover the extrinsic curvature for the neutral case,
and the flat case result, if we further set $M=0$~\footnote{The curvature
is expressed in terms of $\bar r$, because the junction conditions
will be employed below in the reference frame of the areal coordinate,
and not of the isotropic $r$.},
\be
K_{\theta\theta}^0
=
\bar r
\ .
\label{K0}
\ee
\par
The total energy of this spherically symmetric space-time with asymptotically
flat metric (for $\bar r\sim r\to\infty$) is given by the surface integral
\be
E=
-\lim_{R\to\infty}
\left[
\int \frac{\ud\theta\,\ud\phi}{16\,\pi}\sqrt{g^{(2)}}
\left(K-K^0\right)_{\bar r=R}
\right]
\ ,
\label{Eadm}
\ee
where $g^{(2)}$ and $K$ are, respectively, the determinant of the two-metric and the trace
of the extrinsic curvature of a two-sphere of areal radius $R$.
The trace $K^0$, obtained by embedding the two-sphere in a three-dimensional Euclidean
space, yields the Minkowski ``reference'' energy.
It is easy to see that, from Eqs.~\eqref{RN}, \eqref{K} and \eqref{K0}, one obtains 
\be
E
=
M
\ ,
\ee
which shows that $M$ is indeed the ADM mass of the system,
as expected~\cite{adm60}, and regardless of the value of $Q$.
\subsection{Shell energy-momentum tensor}
The shell matter at isotropic $r=\epsilon$ is represented by a $\delta^{(3)}$-function
energy density,
\be
\sqrt{g^{(3)}}\,T^t_{\ t}
=
-\frac{m_0}{2}\,\sqrt{\eta^{(3)}}\,\delta^{(3)}(r)
\ ,
\ee
where $g^{(3)}$ is the determinant of the spatial metric, 
$\eta^{(3)}=r^4\,\sin^2\theta$, and
\be
4\pi\int_0^\infty \delta^{(3)}(r)\,r^2\,\ud r=1
\ ,
\label{d3}
\ee
with $\delta^{(3)}(r)=0$, for $|r-\epsilon|>0$.
We just recall here that $\epsilon\ge 0$ for $|Q|\le M$, whereas $\epsilon\ge r_{\rm min}$
given in Eq.~\eqref{rmin} for $|Q|>M$.
\par
One can more easily describe the shell in comoving coordinates
such that the $(2+1)$-dimensional metric on the shell reads
\be
\ud s^2_{\epsilon}
=
-\ud\tau^2
+R^2\left(\ud\theta^2+\sin^2\theta\,\ud\phi^2\right)
\ ,
\label{shellg}
\ee
and the shell energy momentum tensor is diagonal, 
\be
T_{ij}
=
{\rm diag}\,\left[
-m_0,-R^2\,p,-R^2\,(\sin^2\theta)\,p
\right]
\ ,
\ee
where $p$ is the surface tension.
\subsection{Junction equations}
\begin{figure}
\centering
\raisebox{3.5cm}{$\frac{M}{m_0}$}
\epsfxsize=7cm
\epsfbox{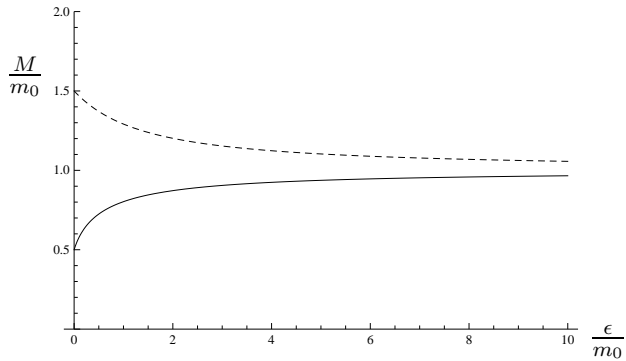}
$\frac{\epsilon}{m_0}$
\caption{ADM mass $M/m_0$ versus the areal coordinate $r=\epsilon/m_0$ of the shell
for fixed $m_0$ and $|Q|=m_0/2<m_0$ (solid line) and $|Q|=3\,m_0/2>m_0$ (dashed line).
\label{Mveps}
}
\end{figure}

We shall enforce the junction equations in the radial areal coordinate frame in which
the inner, outer and shell metrics are respectively given by Eqs.~\eqref{flatg},  \eqref{RN}
and \eqref{shellg}.
Continuity of the metric across the shell, $\ud s^2_{\rm i}=\ud s^2_{\rm o}=\ud s^2_{\epsilon}$
at $r=\epsilon$, then implies
\be
R=\bar r(\epsilon)=\bar r_{\rm i}(\epsilon)
\ .
\ee
Moreover, the discontinuity of the extrinsic curvature at the shell surface,
\be
K_{\theta\theta}(r=\epsilon)
-
K_{\theta\theta}^0(r=\epsilon)
=
-m_0
\ ,
\ee
leads to
\be
M^2+2\,\epsilon\,M-2\,\epsilon\,m_0-Q^2=0
\ ,
\label{eqM}
\ee
from which we obtain the final result
\be
M=-\epsilon+\sqrt{\epsilon^2+2\,\epsilon\,m_0+Q^2}
\ ,
\label{Meps}
\ee
namely an ADM mass that depends on the shell (isotropic) radius $r=\epsilon$.
Note that Eq.~\eqref{Meps} reduces to the analogous expression for $Q=0$
in Refs.~\cite{adm60,cgs},
and has the expected asymptotic behaviour for large $\epsilon\gg m_0$, namely
\be
M
\simeq
m_0
\ .
\label{Masy}
\ee
For intermediate radii, the ADM mass interpolates monotonically between $m_0$ and $|Q|$
(see Fig.~\ref{Mveps}).
\section{Small-volume limit}
\label{secQM}
\setcounter{equation}{0}
In Section~\ref{secCADM}, we shall analyse what happens to the system when the
shell isotropic radius is taken to its mathematically possible lowest value~\cite{adm60}.
We will consider the cases $|Q|\le M$ and $|Q|> M$ separately, so that the minimum value
of $\epsilon$ is either $0$ or $r_{\rm min}$.
Subsequently, in Section~\ref{secQADM}, we shall impose a minimum length
to the shell areal radius and derive physical consequences like in Ref.~\cite{cgs}.
\subsection{Classical model}
\label{secCADM}
\begin{figure}[t!]
\centering
\raisebox{3.5cm}{$\frac{M^2-Q^2}{m_0^2}$}
\epsfxsize=7cm
\epsfbox{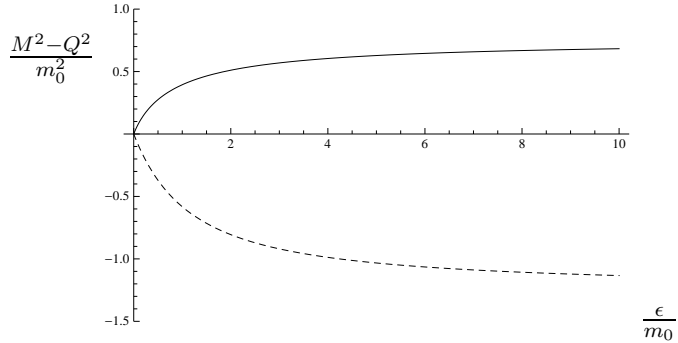}
$\frac{\epsilon}{m_0}$
\caption{Difference $M^2-Q^2$ versus $\epsilon$ for fixed $m_0$ and $|Q|=m_0/2<m_0$ (solid line)
and $|Q|=3\,m_0/2>m_0$ (dashed line).
In the former case, the outer geometry contains a throat of radius~\eqref{R+eps}, whereas in the latter
it does not.
\label{M2-Q2}
}
\end{figure}
\begin{figure}[t!]
\centering
\raisebox{3.5cm}{$\frac{\bar r}{m_0}$}
\epsfxsize=7cm
\epsfbox{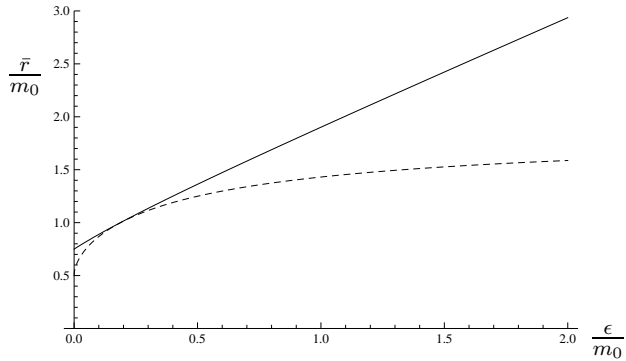}
$\frac{\epsilon}{m_0}$
\caption{Areal coordinate $\bar r=\bar r(\epsilon)$ of the shell (solid line) and
throat radius $\bar R_+(\epsilon)$ (dashed line) for fixed $m_0$ and $|Q|=m_0/2$.
\label{r_eps}
}
\end{figure}
For $|Q|\le M$, one can consider a vanishing isotropic radius, $\epsilon\to 0$,
and Eq.~\eqref{Meps} simply yields the result of Ref.~\cite{adm60}, namely
\be
M=|Q|
\ ,
\label{MQ}
\ee
which shows that the bare mass $m_0$ does not have any gravitational
effects (at large distance).
In fact, one has $M=0$ in the neutral case (see also Ref.~\cite{cgs} for a
detailed analysis).
\par
A vanishing isotropic radius however does not necessarily mean the shell has
reduced to a point.
Due to the dependence of $M$ on $\epsilon$ given in Eq.~\eqref{Meps},
the shell areal radius [given by Eq.~\eqref{barrr} for $r=\epsilon$] becomes
\be
\bar r(\epsilon)
=
\frac{1}{2}
\left(m_0+\epsilon
+\sqrt{\epsilon^2+2\,\epsilon\,m_0+Q^2}
\right)
\ ,
\label{reps}
\ee
and remains finite for vanishing isotropic shell radius $r=\epsilon\to 0$
(see Fig.~\ref{r_eps}),
\be
\bar r(\epsilon\to 0)
=
\frac{m_0+|Q|}{2}
\equiv
\bar r_0
\ ,
\label{reps0}
\ee
This means that the considered configuration is not really point-like in its
naive sense, but has a ``size'' (area) determined by both charge and bare mass.
Note also that the expression for the throat radius $\bar R_+$ from Eq.~\eqref{Rh}
becomes
\be
\bar R_+
=
-\epsilon+\sqrt{\epsilon^2+2\,\epsilon\,m_0+Q^2}
+\sqrt{2\,\epsilon}\,\sqrt{m_0+\epsilon
-\sqrt{\epsilon^2+2\,\epsilon\,m_0+Q^2}}
\ ,
\label{R+eps} 
\ee
which is real only provided $m_0\ge |Q|$, as can be seen from the difference
$M^2(\epsilon)-Q^2$ plotted in Fig.~\ref{M2-Q2}.
Therefore, for $m_0>|Q|$, we find the very important consistency check that
\be
\bar r_0>\bar R_+
\ .
\label{r0>R}
\ee
It is actually easy to see that $\bar r(\epsilon)\ge\bar R_+(\epsilon)$ for all $\epsilon\ge 0$
(see Fig.~\ref{r_eps}), which means that the shell never disappears behind its horizon
(recall that the space-time for $\bar r< \bar r_0$ is actually flat Minkowski), and this
model does not describe (charged) particles as black holes, but as worm-holes.
More precisely, since 
\be
\lim_{\epsilon\to 0}M^2(\epsilon)=Q^2
\ ,
\ee
the outer geometry becomes that of an extremal Reissner-Nordstrom space-time. 
\par
For $|Q|> M$, we can only take $\epsilon\to r_{\rm min}$ [see Eq.~\eqref{rmin}].
From Eq.~\eqref{Meps} and Fig.~\ref{M2-Q2}, we see that this case is generated by $|Q|>m_0$.
However, for the same reason, we also see that $r_{\rm min}=r_{\rm min}(M)$ becomes
\be
r_{\rm min}
=
\frac{1}{2}
\left(\epsilon+|Q|
-\sqrt{\epsilon^2+2\,\epsilon\,m_0+Q^2}
\right)
\ ,
\ee
and vanishes for $\epsilon\to 0$.
This means that it is legitimate to take $\epsilon\to 0$ also for $|Q|>m_0$ and, since
Eq.~\eqref{MQ} still holds, the outer geometry of the final configuration is an
extremal Reissner-Nordstr\"om metric like for $|Q|\le m_0$.
\par
The overall conclusion is therefore that, from the point of view of an observer
placed in the region outside of the charged shell, at $\bar r>\epsilon$,
after the limit $\epsilon\to 0$, the geometry is given by the 
Reissner-Nordstr\"om metric with $M=|Q|$, regardless of whether $|Q|\le m_0$
or $|Q|> m_0$.
Finally, for a neutral particle, the known limit is recovered by simply setting
$Q=0$, which yields $\bar r_0=m_0/2$ and $\bar R_+=0$~\cite{cgs}.
This known result, if applied to elementary particles, naturally raises some questions
about the implementation of the equivalence principle, since the bare shell mass
$m_0$ becomes unobservable in this limiting process. 
However, elementary particles should also be described by quantum mechanics,
which brings us to the next step.
\subsection{Quantum models}
\label{secQADM}
In order to further clarify the above classical results, let us display the Newton constant
$G=\lp/\mpl$ and Planck constant $\hbar=\lp\,\mpl$ explicitly.
This means, for example, that the bare (geometrical) mass
\be
m_0=\lp\,\frac{\mu_0}{\mpl}
\ ,
\ee
where $\mu_0$ has units of mass, and the geometrical charge
\be
|Q|
\simeq
10^{8}\,\lp\left(\frac{\mu_0}{\mpl}\right)\left|\frac{q}{e}\right|
\ .
\ee
For astrophysical objects, the total charge $q$ is negligible and the condition $|Q|\ll m_0$
therefore usually holds.
However, for standard model particles with an electric charge $q$ equal to (a fraction of)
the electron charge $e$, the geometrical charge $|Q|\simeq 10^8\,m_0$,
and the throat radius~\eqref{R+eps} becomes imaginary.
Moreover, since $\mu_0\ll\mpl$, one has
\be
m_0\ll |Q|\ll \lp
\ .
\ee
We have already shown that the classical limit $\epsilon\to 0$ can also be performed
for $|Q|>m_0$, so that the left inequality is no issue.
However, dealing with classical lengths below the Planck size is definitely more questionable.
\par
Let us then repeat the analysis of the neutral case from Ref.~\cite{cgs}, and introduce 
a minimum length scale $\lambda$, such that both the shell areal radius and the
throat radius (when properly defined) cannot be shorter than $\lambda$, 
\be
\bar r(\epsilon)
\ge
\bar r(\epsilon_\lambda)
\equiv
\bar r_\lambda
\simeq
\lambda
\label{r>L}
\ee
and
\be
\bar R_+(\epsilon)
\ge
\bar R_+(\epsilon_\lambda)
\equiv
\bar R_\lambda
\simeq
\lambda
\ .
\ee
From Eq.~\eqref{reps}, we obtain
\be
\epsilon_\lambda
=
\lambda\left(1-\frac{m_0}{2\,\lambda}\right)^2
-\frac{Q^2}{4\,\lambda}
=
\lambda
-m_0
+\frac{m_0^2-Q^2}{4\,\lambda}
\ ,
\ee
and the ADM mass becomes
\be
M_\lambda
=
m_0\left(1-\frac{m_0}{2\,\lambda}\right)
+\frac{Q^2}{2\,\lambda}
=
m_0
+
\frac{Q^2-m_0^2}{2\,\lambda}
\ ,
\label{ML}
\ee
which shows that the asymptotic expression~\eqref{Masy} is recovered provided
$\lambda\gg |Q|$, $m_0$. 
The limit $\epsilon\to 0$ and the classical result~\eqref{MQ} can be recovered
by simply setting $\lambda=\bar r_0$.
Moreover, the analogous expressions for the neutral case in Ref.~\cite{cgs} are
recovered by setting $Q=0$.
\par
Of course, since Eq.~\eqref{r0>R} holds in general, these results are meaningful
only provided $\lambda\gtrsim\bar r_0$ in Eq.~\eqref{reps},
that is
\be
\lambda
\gtrsim
\left\{
\begin{array}{ll}
\strut\displaystyle
|Q|
\simeq
10^{8}\,m_0
\sim
10^{8}\,\lp\,\frac{\mu_0}{\mpl}
&
\quad
{\rm for}\
q\simeq e
\\
\\
\strut\displaystyle
\frac{m_0}{2}
\sim
\lp\,\frac{\mu_0}{\mpl}
&
\quad
{\rm for}\
q=0
\ ,
\end{array}
\right.
\ee
Note that both bounds above are much shorter than the Planck length for standard
model particles, whose mass $\mu_0\lesssim 10^{-16}\,\mpl$.
\par
In a quantum theory, we could therefore assume that
\par
1) $\lambda\simeq \lp$, and $\lambda$ is a truly fundamental length, or
\par
2) $\lambda=\lp\,\mpl/\mu_0\equiv \lambda_\mu$
is the Compton length of the particle, with $\lambda_\mu\gg\lp$.
\par
\noindent
In either case,
\be
m_0\ll |Q|\ll \lambda
\ ,
\label{Lgg}
\ee
and we can therefore approximate
\be
M_\lambda
\simeq
m_0
\ .
\ee
Note this result holds irrespectively of the sign of the difference $|Q|-m_0$,
as long as Eq.~\eqref{Lgg} is valid, and coincides with what was obtained 
for the neutral case in Ref.~\cite{cgs}.
It therefore the effect of the existence of a minimum (quantum mechanical) length
swaps the roles of $Q$ and $m_0$:
the later becomes the observable ADM mass, in complete agreement with the
literal interpretation of the equivalence principle, whereas the former remains just
the source of the electric field.
\section{Conclusions and outlook}
\label{secConc}
\setcounter{equation}{0}
We have reworked the old shell model of elementary particles of Ref.~\cite{adm60}
and shown, in a rather pedagogical manner, why quantum corrections may be
responsible for restoring the classical equivalence principle for massive charged particles,
precisely in the same way it was previously derived for neutral particles in Ref.~\cite{cgs}.
\par
It would be tempting at this point to speculate whether there exists a different
implementation of the limiting procedure $\epsilon\to 0$ and obtain, instead of
a regular geometry, a true black hole space-time.
Of course, one such way is the actual collapse of a homogenous sphere of dust 
that ends with a Schwarzschild black hole (the Oppenheimer-Snyder model~\cite{OS}).
Beside mathematical differences, at the quantum level, one could then
devise a means to investigate transitions between the two different outcomes
of the ``collapse'': regular particles on one side, and black holes on the other.
This would provide a basic building block for describing both the formation of
black holes from the collision of quantum particles~\cite{orlandi},
and the late stages of the inverse process of Hawking evaporation~\cite{hawking}.
One could also try to bridge the classical and quantum descriptions of
such ``point-like states'', by employing a specific model of black holes
of the form recently proposed in Refs.~\cite{gomez,co}.
\end{document}